\documentclass[12pt,preprint]{aastex}
\begin{document}
\title{On the Origin of the Slow Speed Solar Wind: Helium Abundance Variations}


\author{Cara E. Rakowski \& J. Martin Laming}
\affil{Space Science Division, Naval Research Laboratory Code 7674L,
Washington, D.C. 20375}

\begin{abstract}
The First Ionization Potential (FIP) effect is the by now
well known enhancement in abundance over photospheric
values of Fe and other elements with first ionization potential
below about 10 eV observed in the solar corona and slow speed solar wind. In our model, this fractionation is achieved by means of the ponderomotive force, arising
as Alfv\'en waves propagate through or reflect from steep density gradients in the solar
chromosphere. This is also the region where low FIP elements are ionized, and high FIP elements
are largely neutral leading to the fractionation as ions interact with the waves but neutrals do not. Helium, the element with the highest FIP and consequently the last
to remain neutral as one moves upwards can be depleted in such models.
Here, we investigate this depletion for varying loop lengths and magnetic field strengths.

Variations in this depletion arise as the concentration of the
ponderomotive force at the top of the chromosphere varies in response to Alfv\'en wave
frequency with respect to the resonant frequency of the overlying coronal loop, the magnetic field, and possibly also the loop length. We find that stronger depletions of He are obtained for
weaker magnetic field, at frequencies close to or just above the loop resonance. These results may have relevance to observed variations of the slow wind solar He abundance with wind speed, with slower slow speed solar wind having a stronger depletion
of He.
\end{abstract}

\keywords{Sun:abundances -- Sun:chromosphere -- turbulence -- waves}

\section{Introduction}
Our knowledge of the solar wind has come a long way since the original prediction of
\citet{parker58}. It is now known to have two separate components; ``fast wind'' which
originates in coronal holes with speeds generally greater than 600 km s$^{-1}$, and
``slow wind'' with speeds less than 600 km s$^{-1}$, the precise solar origin of which is
uncertain. The fast solar wind is usually rather steady in speed and mass flux, and is
generally taken to be the solar wind of Parker's prediction \citep{antiochos11},
although it is now recognized to be accelerated close to the sun by the damping
of ion cyclotron and MHD waves \citep{ofman10} rather than by heat conduction as in the original concept \citep{parker63}.

By contrast the slow solar wind is unsteady, with variability presumably reflecting
solar source regions subject to varying phenomena associated with magnetic activity on the
Sun. The slow solar wind is also distinguished from the fast by the properties of its
turbulence. The fast wind exhibits essentially purely Alfv\'enic turbulence, whereas the
slow wind also has magnetic structures advected within it \citep[e.g.][]{bruno05}.
Since such structures may also be represented by superpositions of counter-propagating
Alfv\'en waves in equal amplitude, turbulence in the slow wind is sometimes said to be
``more balanced''. The elemental composition of the two types of solar wind is
also known to be different. Solar plasma in closed coronal loops and in the slow speed
wind is known to be subject to a fractionation process, as the material that supplies
these features moves up from the chromosphere to the corona. Elements which are
strongly ionized in the chromosphere like Fe, Si, Mg, with first ionization potentials
less than about 10 eV (low FIP elements), are enhanced in abundance by a factor of about
3-4. High FIP elements are generally unaffected. He often shows a depletion in the solar
corona \citep{laming01,laming03} and wind, which is larger and more variable in the slow wind \citep{aellig01,kasper07} than in the fast \citep{reisenfeld01}. The
abundance enhancement of low FIP ions in the fast solar wind is also much smaller than
in the closed loop corona or the slow wind.

This variability arises naturally in models of the FIP effect \citep[see][and discussed in more detail below]{laming04,laming09,laming11}, where the agent of fractionation is the ponderomotive
force associated with Alfv\'en waves propagating between the chromosphere and the corona.
The changing magnetic field geometry between open and closed field lines alters
the propagation of Alfv\'en waves in such a way that the FIP fractionation produced
by the ponderomotive force varies in the manner observed, i.e. more fractionation on
closed field lines. Thus we assume henceforward that one fundamental difference between
fast and slow solar wind is that slow solar wind must have originated on a closed coronal
loop to become fractionated, and that this loop is subsequently opened up to release its
plasma into wind.

Some authors argue that this distinction is not required. While the slow wind speed is
well known to correlate inversely with the expansion rate of open field in equatorial
coronal holes \citep{wang90} leading to the association of the slow wind with super-radially expanding flux tubes at the edges of coronal holes \citep[e.g.][]{ko06},
the view that the slow wind originates exclusively in open field regions has been
advanced by \citet{cranmer07}. They use their simulated Alfv\'en wave spectra in polar
and equatorial coronal holes to calculate FIP fractionation following \citet{laming04}
and find a difference between fast wind from polar holes and slow wind from equatorial
hole similar to that that is observed. On this basis, the transition in elemental
composition from fast to slow wind should probably be gradual. However in observations
\citep{geiss95,zurbuchen99} this transition is abrupt, both in the change in composition and
the freeze-in temperature associated with the O charge states. \citet{antiochos11} also
comment that a tight correlation between element abundances and wind speed should
result from such models, a finding which also is not supported by observation. We
therefore adopt the view that the fast and slow winds are distinct entities, and
the distinction arises as the fast wind originates exclusively in open field, while to become fractionated, the plasma comprising the slow wind must have spent time inside a closed coronal
magnetic loop, which is subsequently opened up by reconnection.

In this paper we will investigate the abundance pattern expected from such a process.
We follow \citet{laming11} in assuming the Alfv\'en waves to have a coronal origin,
presumably associated with heating mechanisms such as nanoflares and their associated
reconnection \citep[e.g.][]{sturrock99} or resonant absorption \citep[e.g.][]{ruderman02}.
Due to repeated reflection at each loop footpoint \citep{ofman02}, equal intensity Alfv\'en waves
propagating in each direction are expected to be
produced by these processes. We pay particular
attention to variations of the abundance of helium, observed to change with slow solar wind speed
\citep{aellig01,kasper07}. We show that not only does the FIP effect model based on the ponderomotive force \citep{laming04,laming09,laming11} explain the fact that the solar wind abundance of helium is {\em lower} than its generally accepted photospheric value, but that
variations in this abundance may also be within reach of the model. While processes such as gravitational settling or inefficient Coulomb drag are not completely ruled out, their
importance in forming the elemental composition of the solar wind is reduced from
that discussed in previous works \citep{bochsler07,kasper07}.

\section{The Ponderomotive Force Model}
In the model developed by \citet{laming04,laming09,laming11}, FIP fractionation has
its origin in the action of the ponderomotive force on chromospheric ions, but
not neutrals. The ponderomotive force is a second order term incorporating the
$\rho\delta {\bf v}\cdot\nabla\delta {\bf v}$ and  $\delta {\bf
J}\times\delta {\bf B}/c$ terms in the MHD momentum equation. In many cases it looks like
a force due to the gradient of wave magnetic pressure, but this obscures its true
nature as the interaction of waves and particles through the refractive index of the plasma. Given this, it can be seen that both upward and downward ponderomotive
accelerations may occur, with the latter giving a potential explanation for the
so-called ``inverse FIP'' effect seen in active stars and in dwarfs of later spectral
types than the sun \citep[e.g.][]{wood10}. The fractionation arises from the
time averaged part of the ponderomotive
force, which for waves of
frequency $\omega _A<< \Omega$, the ion gyrofrequency, is given by  (e.g Litwin \& Rosner 1998, Laming 2009)
\begin{equation}
F={\partial\over\partial z}\left(q^2\delta E_{\perp}^2\over 4m\Omega ^2\right)
={mc^2\over 4}{\partial\over\partial z}\left(\delta E_{\perp}^2\over B^2\right),
\end{equation}
where $q$ is
the ion charge. The ponderomotive acceleration, $F/m$, is independent of
the ion mass.

Our basic approach builds on \citet{hollweg84}, and is illustrated by a cartoon in Figure 1. We start with a downward propagating
Alfv\'en wave at the $\beta =1.2$ layer of one footpoint of a coronal loop,
and integrate the non-WKB transport
equations spatially for the forward and backward propagating Alfv\'en waves through the corona to the other footpoint, where we now have a mixture of waves propagating in both
directions, depending on the resonant properties of the loop with respect to
the wave frequency. The transport equations are \citep[][and refs therein]{cranmer05}
\begin{equation} {\partial z_{\pm}\over\partial t}+\left(u\pm
V_A\right){\partial z_{\pm}\over\partial r}= \left(u\pm
V_A\right)\left({z_{\pm}\over 4H_D}+{z_{\mp}\over 2H_A}\right),
\end{equation}
where $z_{\pm}=\delta v_{\perp}\pm \delta B_{\perp}/\sqrt{4\pi\rho }$ are the
Els\"asser variables for Alfv\'en waves propagating against or along
the magnetic field respectively. The Alfv\'en speed is $V_A=B/\sqrt{4\pi\rho}$, the
upward flow speed is $u$, the mass density is $\rho $, and the magnetic field is $B$.
The signed scale
heights are $H_D=\rho /\left(\partial\rho /\partial r\right)$ and
$H_A=V_A/\left(\partial V_A/\partial r\right)$. In the solar
chromosphere and corona $u << V_A$, and we put $u=0$. Once $z_{\pm}$ are determined by numerical integration of equation 2, we
calculate $\delta v_{\perp}$ and $\delta B_{\perp}/\sqrt{4\pi\rho}$, and hence the
perpendicular electric field $\delta E_{\perp}$ from
\begin{eqnarray}
\delta v_{\perp}&={z_++z_-\over 2}\cr
{\delta B_{\perp}\over\sqrt{4\pi\rho}}&={z_+-z_-\over 2}\cr
\delta E_{\perp}&={\delta v_{\perp}B\over c}={\delta B_{\perp}V_A\over c}.
\end{eqnarray}
Contributions from waves of the same frequency propagating in both directions along the magnetic
field are naturally accounted for. In this work, for reasons of pedagogy, we only consider one
pair of parallel forward and backward propagating Alfv\'en waves of a given
frequency at a time, and neglect any nonlinear effects or dissipation. We are tacitly assuming
that loop dynamics are mainly driven Alfv\'en wave interactions, as
suggested by \citet{bigot08}.

The degree of fractionation is given by the formula
\begin{equation}
{\rm fractionation} =\exp\left(2\int
_{z_l}^{z_u}{\xi _sa(\nu _{eff}/\nu _{s,i})/v_s^2}dz\right)
\end{equation}
(see
equation 9, Laming 2009, equation 12, Laming 2004, which follow
Schwadron et al. 1999), where $\xi _s$ is the ionization fraction of
element $s$, $a$ is the ponderomotive acceleration, $\nu _{eff}=\nu
_{s,i}\nu _{s,n}/\left[\xi _s\nu _{s,n}+\left(1-\xi _s\right)\nu
_{s,i}\right]$ where $\nu _{s,i}$ and $\nu _{s,n}$ are the collision
rates of ions and neutrals, respectively, of element $s$ with the
ambient gas. Also $v_s^2 =kT/m_s +v_{\mu turb}^2 +v_{turb}^2$, where
$v_{\mu turb}$ is the amplitude of microturbulence coming from the
chromospheric model, and $v_{turb}$ is the amplitude of longitudinal waves induced
by the Alfv\'en waves themselves. These derive from the harmonic variation of the
ponderomotive force, and unlike the fractionation, do not require an external density
gradient. A full description of this is given in \citet{laming11}.
The limits of integration, $z_l$ and $z_u$ bracket the
region where the ponderomotive acceleration is effective in the chromosphere. Typically
$z_l$ is at the $\beta = 1.2$ layer in the low chromosphere, and $z_u$ is in the transition
region where all elements are ionized.

The chromosphere at each footpoint can be based on any of the \citet{vernazza81}
models or similar. Here we use the \citet{avrett08} update of
VALC. The ionization balance of the minor ions is computed at each
height in the chromosphere using the model temperature and electron
density, and a coronal UV-X-ray spectrum appropriately absorbed in
the intervening chromospheric layers. \citet{laming11} investigated different
approximations for computing the chromospheric ionization balance, and while the
degree of FIP fractionation varied somewhat, the validity of this explanation
for the phenomenon remained intact. The chromospheric magnetic field (illustrated in
Figure 2) is taken to
be a 2D force free field from Athay (1981) and designed to match chromospheric
magnetic fields in Gary (2001) and Campos \& Mendes (1995), which represents the
expansion of the field from the high $\beta$ photosphere where the field is
concentrated into small network segments, into the low $\beta$ chromosphere where the
field expands to fill much more of the volume. The coronal section of the loop has constant magnetic field, and a density scale height of 125,000 km.

The precise model for the chromospheric magnetic field is not critical, because most of the
FIP fractionation occurs in the upper chromosphere where the field is most uniform and vertical.
In the lower chromospheric
region with curving magnetic field in the plasma layer where sound and Alfv\'en speeds converge,
contrary to the case in the corona, wave interaction processes play an important role. We do
not attempt to model these directly, but argue that the increased longitudinal wave amplitudes
in this region, arising from mode and parametric conversion, increase the value of $v_s^2$ in the
denominator of the integrand in equation 4 to such an extent so as to render fractionation
insignificant in this region. In \citet{laming11}, the longitudinal and Alfv\'en wave intensities
are coupled by equation 15 in that reference, motivated by calculations illustrated in \citet{cranmer07} and \citet{khomenko10}. Here we note that recent work by \citet{heggland11} suggests that sound waves may be strong in this region even without the presence of Alfv\'en waves, as a result of turbulent motion in the convection zone.

\section{A Parameter Survey of FIP Fractionation}

In this paper we begin to explore the parameter space of magnetic field and
loop length looking at coronal magnetic fields of 5G, 10G, 15G, and loop lengths of 
50,000 km, 75,000 km, and 100,000 km.
\citet{laming11} investigated the variation of FIP fractionation in a 100,000 km loop
with 20G magnetic field, giving a loop resonance angular frequency of 0.07 rad s$^{-1}$,
as the Alfv\'en wave driver of the FIP effect varied with respect to this resonance.
With wave amplitudes chosen to give a fractionation Fe/O of about 4 as observed for example
by \citet{bryans09}, on resonance,
He/O was about 0.6, S/O about 1.7, and C/O about 1.2. As the wave frequency increased above the
resonant value, as might happen for example if a coronal heating process excited
waves at the loop resonance, followed by chromospheric evaporation reducing this
resonant frequency, He/O decreased slowly at first and then increased up
towards photospheric values,
S/O increased rapidly then levels off to a maximum of $\sim 1.9$, and C/O
increased rapidly to a maximum of 1.6 and then began decreasing. Other fractionations
varied also, but these three appeared the most robust and easiest to interpret
in terms of available observational data.

Performing similar calculations for a wider variety of loop models, again keeping the Fe/O fractionation to a value of about 4,
we see a similar pattern with Alfv\'en wave frequency as was
seen for 20G and 100,000 km.  In Figure 3 we plot an example of how
the fractionation varies with Alfv\'en wave frequency, for a field of 15 G and
a loop length of 75,000 km.
Overlaid on the plot are the fractionations measured in the \citet{vonsteiger00}
Ulysses survey of the solar wind.
The color bar gives the range of
frequencies examined, which start from just below resonance at 0.651 rad s$^{-1}$.  The
black line is below resonance. The purple line is the resonant
case. The two blue lines that still have strong helium depletion are
just above resonance. The green, yellow and red lines are further from
resonance, have less and less helium depletion and enhanced S and C. In general
the modeled fractionation pattern between Mg, Fe, and Si does not match the
observations, with Fe predicted to have more FIP fractionation than
actually observed in the solar wind.
However other spectroscopic observations \citep{bryans09} actually match this pattern very well
\citep[see][]{laming11}.

In Figure 4, we plot the He/O abundance ratio as a function of frequency, in units of the
loop resonant frequency, for a variety of loop lengths and magnetic field strengths.
The curve for 100,00 km loop length and 20 G magnetic field is very similar to \citet{laming11},
with a minimum He/O fractionation of 0.6 at resonance. The position of this minimum shows a
clear trend to lower He/O and higher frequency above resonance for
weaker magnetic fields (and possibly also longer loop lengths); cases
where the ponderomotive force is more concentrated at the top of the
chromosphere. Figure 5 gives the coronal amplitude of the
Alfv\'en wave in each case.
For 5G and 100,000 km, and 10G with loop lengths 75,000 and 100,000
km it was not always possible to achieve a fractionation for Fe/O of
about 4 for any coronal wave amplitude, so these points were omitted
in Figures 4 and 5 and in Figures 9 and 10 to follow below.
The coronal wave amplitudes at resonance for each case are very similar, suggesting that if we
normalized by this parameter instead of forcing Fe/O $\sim 4$, we would have very similar results.
The curve for 15 G, 75,000 km (the upper green curve) gives the coronal wave amplitudes for
the fractionations shown in Figure 2.

Figure 6 shows the energy transmission characteristics of the various loops considered here, calculated from the fluxes of upgoing and downgoing waves in the loop footpoint encountered at the end of the numerical integration, and displayed with the same color coding as in preceding figures.
On resonance, all loops have an energy transmission of unity. The width of this resonance
increases with decreasing loop length, as does the minimum value of the energy transmission in
between resonances. A trend with loop magnetic field is less easy to discern.

Figure 7 depicts the main features of the simulation for the case
of 15 G, 75,000 km and 0.07 rad s$^{-1}$ just above the resonant
frequency of 0.0651 rad $^{-1}$. Top left gives the density and temperature
structure of the chromosphere to be read on the left (black) and right hand (gray)
axes respectively.
Bottom left gives the ponderomotive acceleration and slow mode wave amplitude. The ``spike''
in the ponderomotive acceleration can be seen to coincide with the steep density gradient
at the top of the chromosphere.
The ionization structure in the chromosphere, illustrated in the top right panel for Fe,
Mg, S, and C, and also in the bottom right panel (gray lines, to be read on the right hand side
axis) along with the FIP fractionations (black lines, to be read on the left axis),
determines which elements can be accelerated and therefore fractionated.

Fe, Mg, S and C are all mostly ionized
throughout the relevant region of the chromosphere, H
and O become ionized between 1500 and 2000 km altitude in the chromosphere, and
He is only ionized at the very top of the chromosphere. With our normalization of
Fe/O $\sim 4$, He becomes depleted relative to O when the ponderomotive acceleration is
concentrated at the top of the chromosphere, i.e. when the ``spike'' associated with the
steep density gradient is very strong, as in the case illustrated. O is accelerated up
into the chromosphere while He is left behind, being the last element in the chromosphere
to remain neutral.
When waves penetrate deeper in the chromosphere Fe, Mg, S and C are all easily
fractionated, and the relative contribution from the top of the
chromosphere where O is accelerated and He left behind is less
important. Since we are modeling specifically for Fe/O of 4, He/O will
be lowest when the acceleration region is confined to the top of the
chromosphere, which happens close to the loop resonant frequency.

There are several factors that play into how deep waves penetrate into
the chromosphere.
Longer wavelength waves resonant with longer
loops are not expected to penetrate so far into the chromospheric density gradients before being reflected
back into the loop. This is clear from the fact that for a given loop,
the first harmonic frequency is slightly less than twice that of the
fundamental. Similar results are known for photospheric Alfv\'en waves launched
upwards \citep{cranmer05,verdini07}. However in our case, this effect is probably rather small, and not discernable on Figure 4. Figure 8 illustrates this in more detail. The four panels show
at top left the real (black) and imaginary (gray) parts of $\delta v_{\perp}$ (dashed lines) and
$\delta B_{\perp}/\sqrt{4\pi\rho }$ (solid lines), at bottom left the upgoing and downgoing wave energy fluxes,
at top right the profile of the Alfv\'en speed, and at bottom right the profiles of
$\partial V_A/\partial z$ (dashed lines) and $\sqrt{\left|V_A\partial ^2V_A/\partial z^2\right|}$
(dash-dot line), again all for
a wave of angular frequency 0.07 rad s$^{-1}$ (horizontal dotted line in the bottom right panel)
on a coronal loop with length 75,000 km and magnetic field 15 G. Expressions for cut-off frequencies, at which the Alfv\'en wave becomes evanescent, differ among various authors, but are of similar
magnitude to quantities like $\partial V_A/\partial z$ and $\sqrt{\left|V_A\partial ^2V_A/\partial z^2\right|}$.
Also plotted as a solid line is $\sqrt{\left(\partial V_A/\partial z\right)^2/4+
\left|V_A\partial ^2V_A/\partial z^2\right|/2}$, which is given as the cutoff frequency by
\citet{musielak92} and \citet{moore91}. As can be seen, most wave reflection occuring at the
steep chromospheric density gradient between 2100 and 2200 km altitude will not be greatly affected
by changing of the Alfv\'en wave frequency in the range 0.01 - 0.1 rad s$^{-1}$. Also, an
increased wave frequency due to increased coronal magnetic field will not change its reflection
properties, because the critical frequency also increases with $V_A$.

So while the energy transmission depends primarily on the loop length, and hence wavelength of
the Alfv\'en wave, from Figure 4, it appears that the magnetic field is more important in
controlling the degree of fractionation. Weaker magnetic field means that the
altitude where the plasma $\beta \sim 1.2$ is higher, and the depth of chromosphere beneath
a loop footpoint where fractionation may take place is reduced.
The slow mode wave generation is also stronger in weaker magnetic
fields, and so as the magnetic field becomes weaker, the more
ponderomotive acceleration can be saturated by the associated
longitudinal pressure.
Additionally, compared to waves precisely on resonance,
waves slightly off-resonance give a very similar ponderomotive acceleration
profile, but generate more slow mode waves in lower regions of the chromosphere, again biasing
fractionation further towards the top of the chromosphere. Thus we may expect the minimum He
abundance to be just to higher frequency than the actual loop resonant frequency, as observed
in our calculations.
Further from resonance the spike in the ponderomotive force generates
both positive and negative accelerations counteracting one another, such that essentially all fractionation
occurs low down in the chromosphere, and He is thus not strongly
depleted.

The importance in this example of the relative suppression of fractionation low in the chromosphere suggests that elements such as C or S might ``echo'' this behavior of He. As explained in
\citet{laming11}, these elements can be fractionated in the low chromosphere, assuming the ponderomotive acceleration is present here, because H is dominantly neutral. This means that
elements fractionate roughly in proportion to their ion fraction, and both C and S are
approximately 90\% ionized. Once H is ionized, fractionation is suppressed unless the charge
state fraction is much closer to 1; for example Fe and Mg have ionization fractions of 0.9995 and
0.998 respectively. Thus when the ponderomotive acceleration is concentrated at the top of the
chromosphere, and when He is strongly depleted, S and C should have relatively little
fractionation, and correspondingly when the acceleration is present in the low chromosphere and
He/O is relatively unaffected, S and C should display some fractionation.

Figures 9 and 10 show C/O and S/O as a function of frequency. Both show
the expected qualitative behavior with frequency and magnetic field, although the minimum
in these ratios appears always to be at the loop resonance, as
a steeply peaked dip. A small
distance above resonance the fractionation is a function of magnetic
field and loop length with the highest fractionations for the highest
magnetic field and the shorter loop lengths. The fractionation peaks
at some value and then declines far away from resonance.

\section{Variation of the FIP Effect with Solar Activity?}
\subsection{The Helium Abundances}
Aellig et al. (2001) and Kasper et al. (2007) both found that the
solar wind Helium abundance varied with wind speed and the solar
cycle. Although there is a large scatter in the individual data points
(Aellig et al. 2001) the He/H ratio averaged over a velocity bin
increases almost monotonically with wind speed. The variation with
wind speed is wider during solar minimum than during solar maximum in
the sense that the strongest average He depletion is seen at solar
minimum.

We calculate fractionations relative to O, on the assumption that all elements
can be treated as minor ions (this is not true for H).
To be certain that this did not compromise our results we
looked at the publicly available level 2 ACE SWICS He/O and He velocity data over the life of the
mission so far. Figure 11
reproduces the findings of Aellig et al. (2001) and Kasper et al. (2007) with respect
to wind speed, in
that lower He velocities correspond to lower average He/O ratios in a
mostly monotonic fashion. We do not find the strong dependence of the helium abundance
on phase of the solar cycle. It must be remembered that we are considering He/O, whereas
\citet{aellig01} and \citet{kasper07} worked with He/H.

These authors followed several antecedents and discussed their results in terms of
inefficient Coulomb drag or gravitational settling. We are going to argue that this
abundance anomaly is part of the same process that gives rise to the FIP effect, but before
doing so we briefly review this alternative. Gravitational settling is known to
occur on the Sun in quiescent streamers \citep{feldman98}, where Fe is observed to be
depleted relative to other elements. \citet{noci97}, following \citet{geiss70},
make the point that if such processes
were the origin of the solar wind He abundance depletion, other heavy ions should also be depleted,
which is now known not to be the case. \citet{bochsler07} revisits this, comparing He, O, and
Ne for which He appears most affected by gravitational settling, or inefficient Coulomb drag, in
the corona. In this case the He abundance should correlate with the proton flux, which is
not observed \citep{wang08}. He and other minor ions are also known to flow
{\em faster} than H in the solar wind observed {\it in situ} at 1 AU
\citep[e.g.][]{neugebauer96,kasper08,bourouaine11} which conflicts with the notion of a
proton driven wind dragging the rest of the minor ions out with it, unless a second stage of
wind acceleration is invoked further out in the heliosphere \citep{wang08}. \citet{laming03}
also measure a depleted He abundance in coronal hole and quiet solar region at altitudes
{\em below} that where He is presumed to be depleted in \citet{bochsler07}.

We believe we can better understand the variation of the solar wind
He abundance in the context of our model of the FIP effect.
Figure 4 shows the variation of He/O as a
function of the frequency relative to the resonant frequency. The
strongest He depletions are only seen just above the resonant
frequency. He/O approaches its photospheric value as one moves further away from
resonance. Consider a loop being heated by a process that also generates Alfv\'en waves.
So long as the perturbation to the magnetic field is small compared to the loop
initial magnetic field, only the
resonant frequencies of the loop would be excited. However the heating
would evaporate material from the footpoints of the loop up into the
loop, changing the resonant frequency. Low levels of activity and
heating would only change the resonance by a small amount. More active
regions would have greater heating, greater evaporation and thus move
the loop further away from the original resonance. The He depletion due
to the originally excited waves then depends on how far from resonance
the loop became. Thus according to our parameter search, higher slow wind speeds
should be associated with plasma that was originally in (shorter) loops with higher
magnetic field.

\subsection{The Sulfur and Carbon Abundances}

In the above picture of loop excitation, we always expect the excited
wave to be slightly off resonance due to chromospheric
evaporation. Thus from figures 9 and 10 we expect the S and C
fractionations to be a function primarily of the magnetic field in the
loop. This means we predict that one should see the average C and S
varying  with solar cycle and wind speed, just as is seen for
He. While S is not available in the level 2 ACE products we can test
this hypothesis for C. In figure 12, we plot the average C/O relative to solar
\citep[0.550][]{asplund09} for 9 velocity bins as a function of time, in the same way as figure
11 for He/O. For at least the years 2000 to 2004 and during 2005 a variation with wind
speed is apparent, with the slowest wind speeds showing a lower C/O
fractionation. The period around 2003 also corresponds to a period of strong He/O
fractionation with wind speed. This fits well with the prediction of our
model. Further confirmation could be found if S/O could be analyzed as
well.
A similar correlation between He/O and C/O has been found \citep[see their Figure 11]{ko06},
who analyzed ACE Level 2 data collected over a week timescale as fast wind from an equatorial
coronal hole transitions to slow wind from the surrounding corona.

\subsection{The Coronal Origin of Alfv\'en Waves}
Our finding that Alfv\'en waves on resonance with a coronal loop give FIP fractionations
in better agreement with observations than other frequencies reinforces our suggestion of
a coronal origin for the waves. No chromospheric origin would naturally achieve such selectivity,
unlike the expectation from a coronal origin. 
For example, consider a flux tube with cross-B density gradient scale length
$l$, undergoing a kink mode oscillation which become Alfv\'enic at a
resonant surface where the wave frequency $\omega = 2L/V_A$, where $L$
is the loop length and $V_A$ is the Alfv\'en speed \citep[see
  e.g.][]{ruderman02}.  The width of this resonant layer is $\delta
\sim \left(l\nu /\omega\right)^{1/3}$, where $\nu = aV_A/R$ is the
kinematic viscosity, $a$ is the loop radius and $R$ is the Reynolds
number. The Alfv\'enic velocity fluctuations are larger than those of
the kink mode by a factor $l/\delta =\left(l^2R/aL\right)^{1/3}\sim
0.1 R^{1/3}$ under typical conditions, and are constrained to be resonant with the flux tube
and their location of excitation.

The nanoflare paradigm suggests that the dominant
loop footpoint motions are of much lower frequency, and do not
excite oscillations in the loop but act so as to build up magnetic
stresses in the corona. These stresses periodically release
themselves, in what has become known as a ``nanoflare'', as a
current sheet develops. Rappazzo et al. (2007, 2008) discuss the
buildup of magnetic stresses within the framework of turbulence
phenomenology, where it appears that velocity perturbations similar
to the $\sim 30$ km s$^{-1}$ observed
should be expected. \citet{longcope09} conjecture that
in impulsive reconnection in post-flare loops, only about 10\% of
the liberated magnetic energy is converted directly into heat, the rest
reappearing as kinetic energy that ultimately drives turbulence. So long as the
perturbation to the magnetic field is sufficiently smaller than the ambient magnetic
field, the boundary conditions imposed by the loop footpoints will restrict the modes
excited by the reconnection event to those that are resonant with the loop.

\citet{sturrock99} gives a pedagogic
review of the mechanisms by which various wave modes may be excited by
reconnection. The reconnected
field line is generally distorted, and this can either propagate away
from the reconnection site as an Alfv\'en wave, or emit
magnetoacoustic waves traveling perpendicularly to the magnetic field
direction. \citet{kigure10} explicitly
consider the generation of Alfv\'en waves by magnetic reconnection,
and find that a significant fraction of the magnetic energy released
(several tens of \%, depending on geometry and plasma $\beta$) can be
carried off by Alfv\'en or magnetoacoustic (fast or slow mode) waves,
with Alfv\'en waves dominating for $\beta <1$. \citet{liu11} discuss the role
of temperature anisotropies and wave generation by the firehose instability
in the outflow.

\section{Origin of Slow Speed Solar Wind}

Based on the variation on the helium abundance in the slow speed wind wind, we argue
that the faster slow speed wind should have originated in magnetic flux tubes with
higher magnetic field, with the slower components of the slow wind arising from
flux tubes with weaker magnetic field. Slow wind speed is known to correlate with the
expansion rate of the open field \citep{wang90}. \citet{wang03} argue that at sunspot maximum,
more slower speed slow wind is produced, and that this comes mainly from small sheared open field regions located near active regions. These have relatively high footpoint field strengths, and
correspondingly high expansion factors. The high coronal heating rate near the footpoint (heating
is presumed to correlate with magnetic field strength) increases the solar wind density,
thereby reducing its speed for the same energy input.
Our inference concerns not this open field considered by \citet{wang08} and \citet{bochsler07},
but the closed field with which it presumably reconnects releasing plasma into the solar wind.

Attempts
have been made to map {\it in situ} slow solar wind observations by Ulysses and ACE spacecraft
back to a magnetic field source surface at a
heliocentric distance of around 2.5 $R_{\sun}$, and thence back to the solar disk itself
\citep{neugebauer02,liewer04}, with the result that at sunspot maximum (but not at activity
minimum), open field lines in active regions are the most likely source.
\citet{wang08} studied the solar wind helium abundance and found that it correlated best with
the source-surface magnetic field strength, which was found to be strong in fast wind and slow wind at sunspot maximum, but weak in the low-speed solar wind emanating from coronal hole
boundaries at solar minimum. Increased proton flux in the low-speed solar wind should
increase the He abundance if inefficient Coulomb drag is important, but as mentioned above,
\citet{wang08} found essentially no correlation between these variables.

If the magnetic field in the originally closed loop correlates
with the strength of neighboring open field with which it must reconnect, our conclusions
match these results. From Figure 3, the dependence of He/O on magnetic field for a given loop
length appears to be much stronger than the dependence on loop length for a given magnetic field,
although there is a tendency for longer loops to have stronger helium depletions.

Our results strongly suggest that the solar wind helium abundance is
set in the chromosphere, along with the other abundance modifications
that result in the FIP effect. \citet{byhring11} reaches a similar conclusion, but argues that it
must arise from gravitational settling. However such processes; diffusion, gravitational settling
and inefficient Coulomb drag are too slow to have an appreciable effect in
a solar atmosphere that is increasingly being appreciated as a dynamic
environment. For example timescales in diffusion models range from tens of hours
\citep{hansteen94} to days or weeks \citep{killie07}, and so such possibilities must be
considered highly implausible.

Our calculations necessarily have approximations in the treatment of wave physics
\citep[see][for fuller discussion]{laming11}. However all of the phenomena we discuss have
qualitative explanations in terms of the structure of the chromosphere and its ionization
balance. While absolute numbers for fractionations may vary slightly in future works as
chromospheric models and atomic physics improve, we expect that the various correlations
between element abundances and the relation of the FIP fractionation to the wave frequency
in terms of the loop resonant frequency to be robust. Most of the variations we have
discussed center around the ``spike'' in the ponderomotive acceleration at the top of the
chromosphere where the density gradient is steep. At first sight it might seem a coincidence
that this ``spike'' occurs where the hydrogen in the chromosphere is becoming ionized, and thus
allows for FIP fractionation. This is not the case. The steep density gradient arises
precisely because hydrogen is becoming ionized, and so cooling by radiation in Lyman $\alpha$ is
becoming quenched. Consequently heat input into the plasma cannot be radiated away, and the
temperature must rise. The density decreases to maintain constant pressure, and waves in this
region give rise to the strong ponderomotive acceleration.

\acknowledgements
This work was supported by NASA Contracts NNH10A055I, NNH11AQ23I, and by basic research funds of
the Office of Naval Research. We are grateful to Dan Moses for a careful reading of the paper.

\clearpage

\begin{figure}
\epsscale{0.75} \plotone{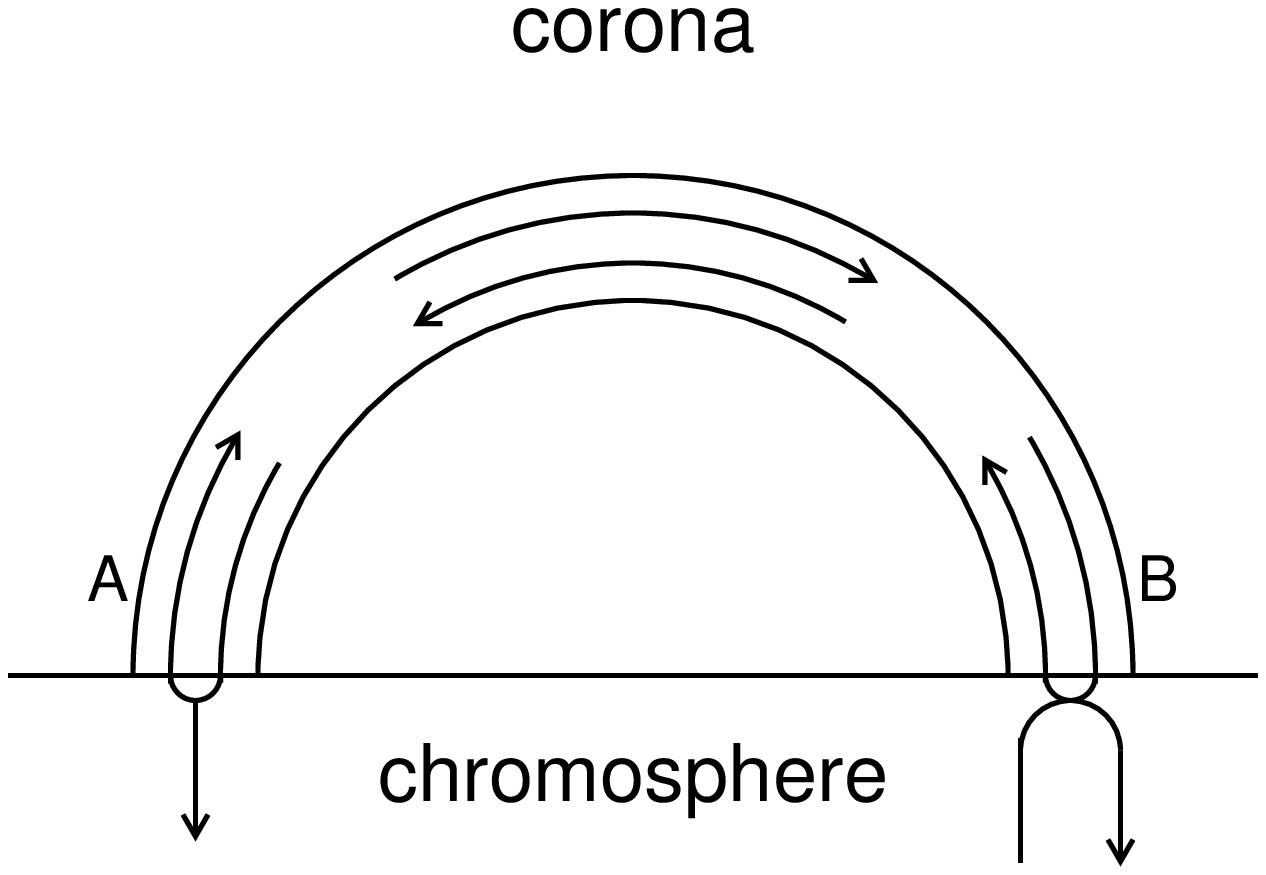} \figcaption[f1.eps]{Cartoon
illustrating the model. The integration of the Alfv\'en wave transport equations is begun
in chromosphere A with one downgoing wave. Integrating back through the corona to chromosphere
B, forward and backward propagating waves develop. At chromosphere B, the wave is either
reflected back into the corona, or transmitted into the chromosphere, depending on the wave
frequency with respect to the resonant frequency of the loop.
Resonant waves reflect, nonresonant waves precipitate down.
\label{fig1}}
\end{figure}
\clearpage

\begin{figure}
\epsscale{0.75} \plotone{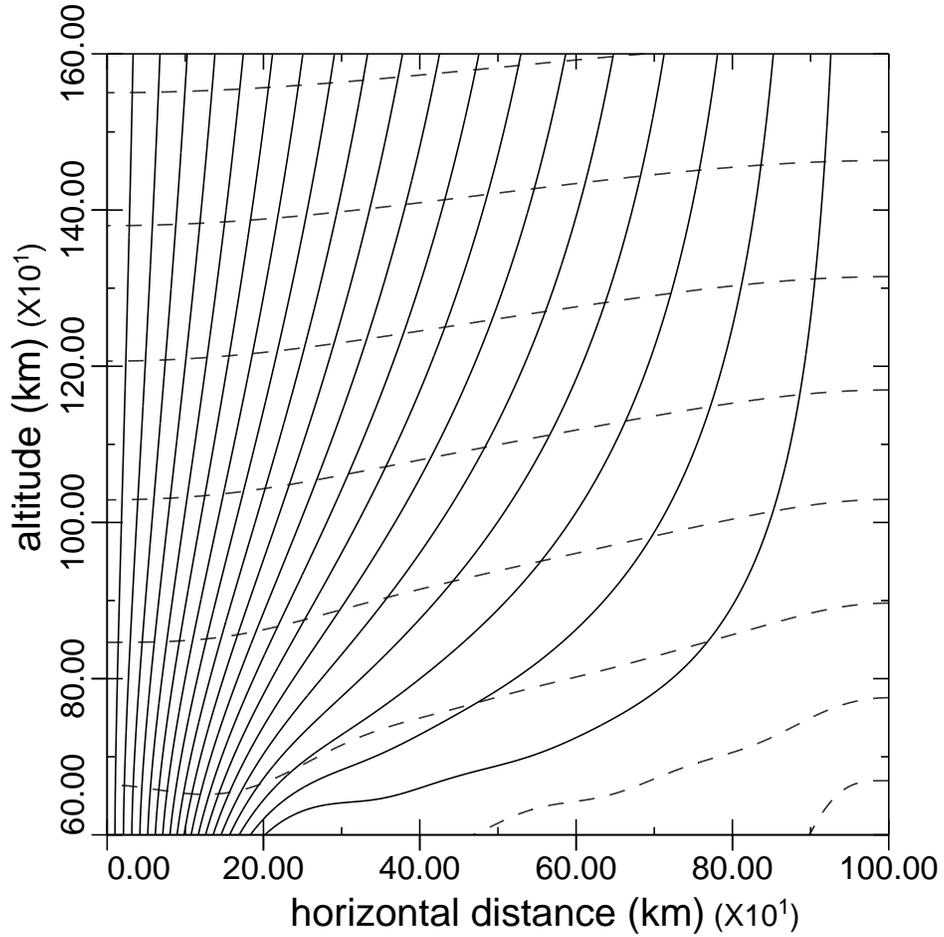} \figcaption[f2.eps]{Depiction of force-free chromospheric magnetic field for coronal loop model with $B = 15$G, calculated from Athay (1981). Solid lines indicate magnetic field lines. Dashed line indicate contours of the Alfv\'en speed. The plasma $\beta = 1.2$ layer is at 700 km altitude above the photosphere, and moves up and down for weaker or stronger coronal fields respectively.
\label{fig2}}
\end{figure}
\clearpage

\begin{figure}
\includegraphics[width=6.4in]{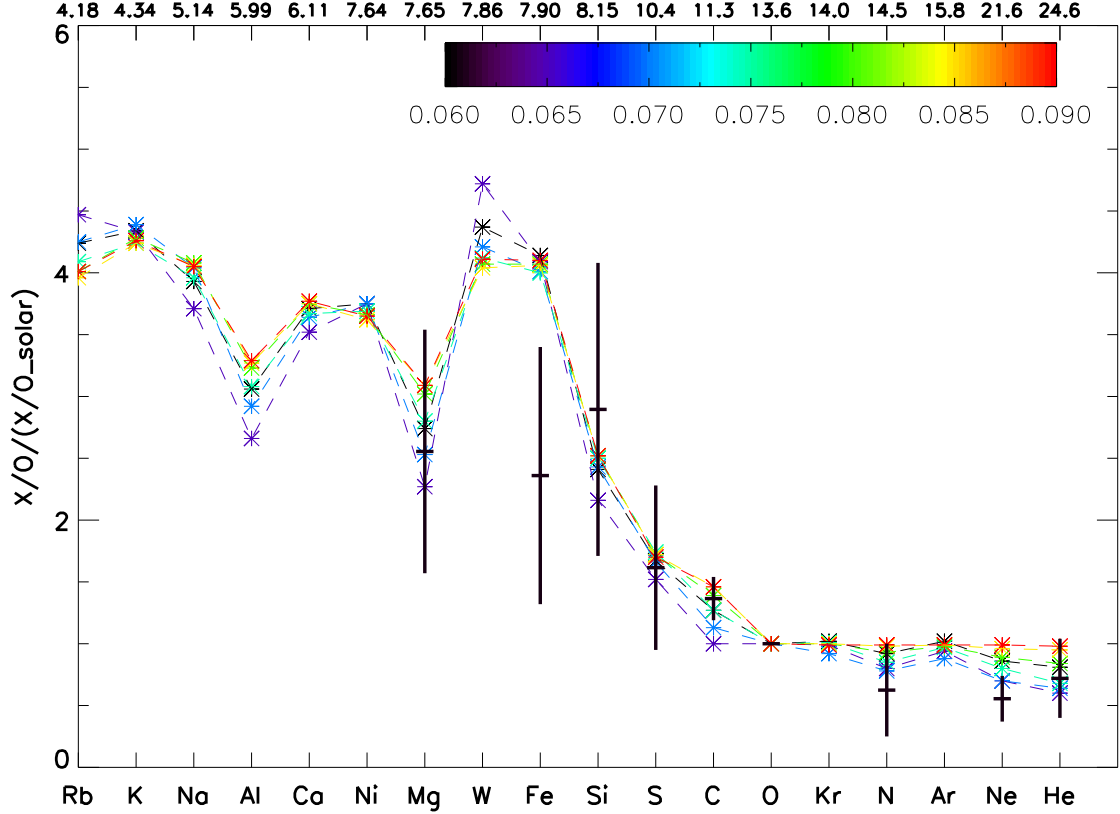}
\caption{Fractionations relative to oxygen relative to their
photospheric values as a function of FIP for the case of 15 G and
75,000 km.  Overlaid are the measured fractionations of the elements included
in von Steiger et al. 2000. The strongest He depletion corresponds to the purple line,
illustrating the resonant case. As one moves off resonance, the He abundance increases. Other
elements are predicted to vary either in correlation (e.g. Al, Mg, S, C), or anti-correlation
(e.g. Rb, W) with He.}
\end{figure}
\clearpage

\begin{figure}
\includegraphics[width=6.4in]{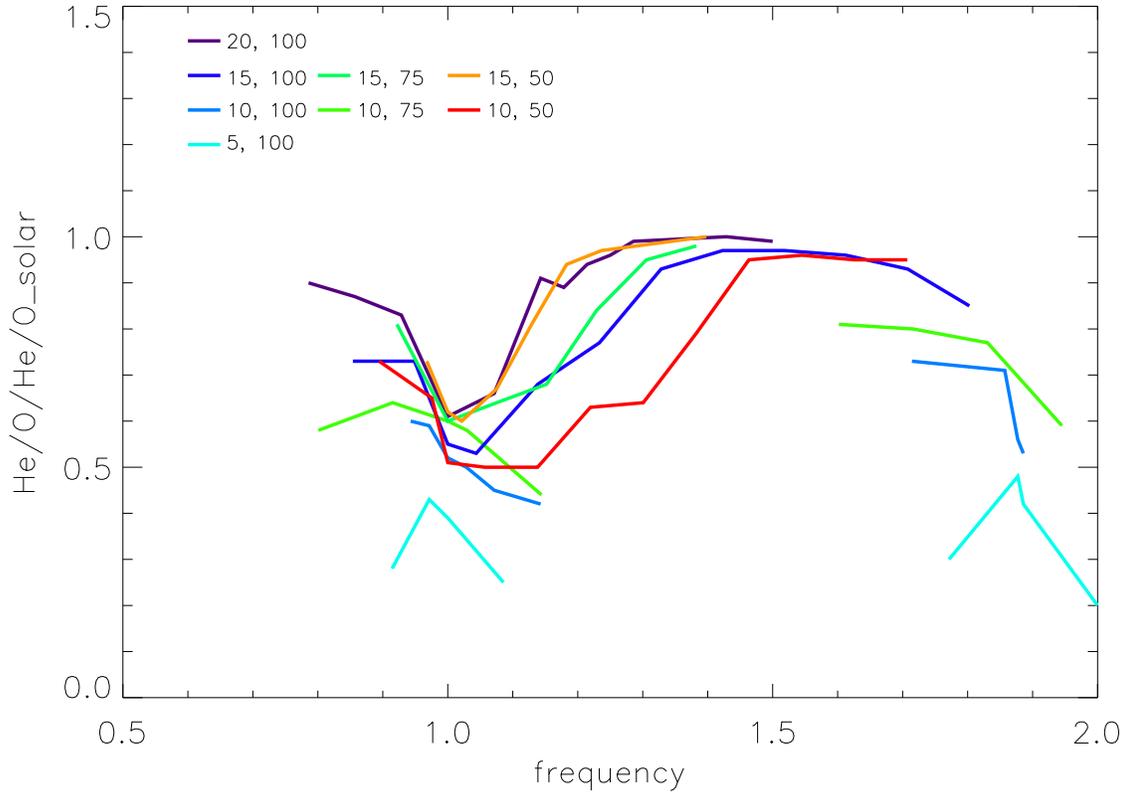}
\caption{He/O relative to their solar ratio as a function of the
frequency divided by the resonance frequency.  For 5G and 100,000 km the
points where it was not possible to obtain Fe/O of 4 for any
amplitude, and these point are omitted. For 10 G and 75,000 or 100,000
km, it was also not possible to achieve Fe/O $\sim 4$ for any amplitude, and these points are
also omitted. A trend of increasing He depletion with decreasing
magnetic field and possibly also increasing loop length can be seen.}
\end{figure}
\clearpage

\begin{figure}
\includegraphics[width=6.4in]{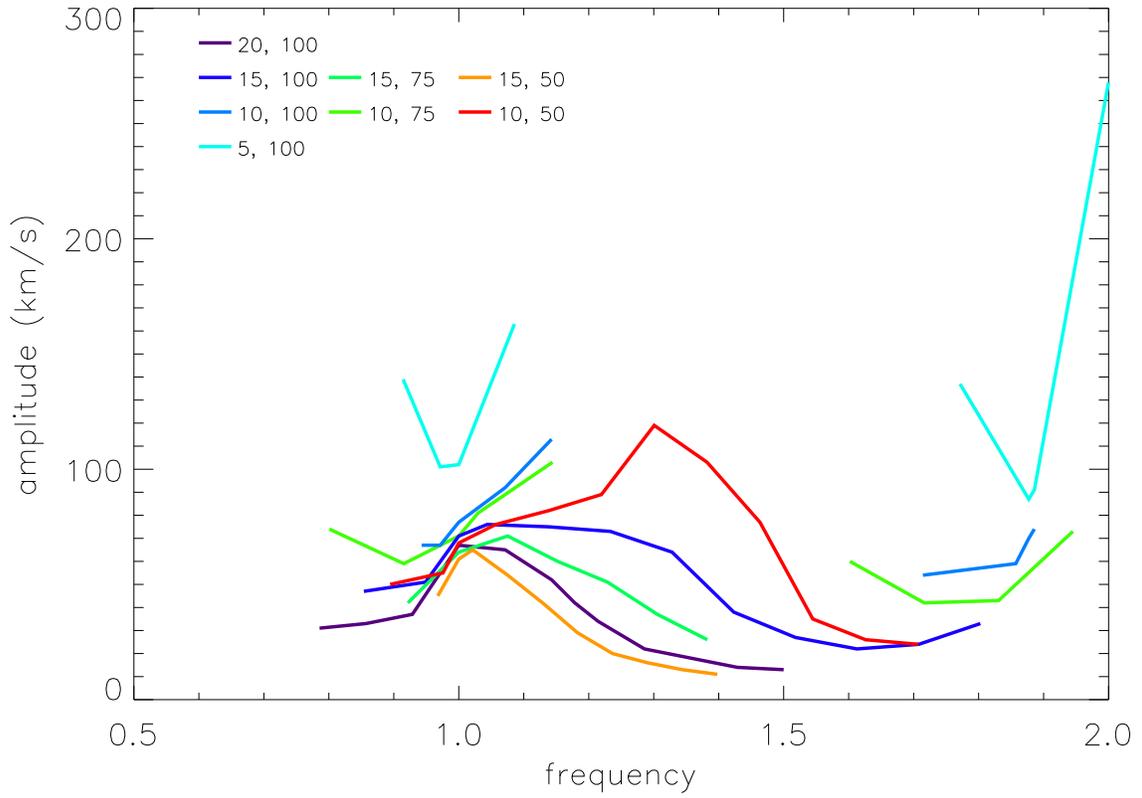}
\caption{Amplitude in km s$^{-1}$ at the peak of the loop as a
function of frequency relative to the resonance frequency. The absent points have
the same meaning as in Figure 3. All curves are selected to give Fe/O close to 4, but as can be
seen, close to resonance, this results in very similar wave amplitudes, even though loop lengths
and background magnetic fields are varying.}
\end{figure}
\clearpage

\begin{figure}
\includegraphics[width=6.4in]{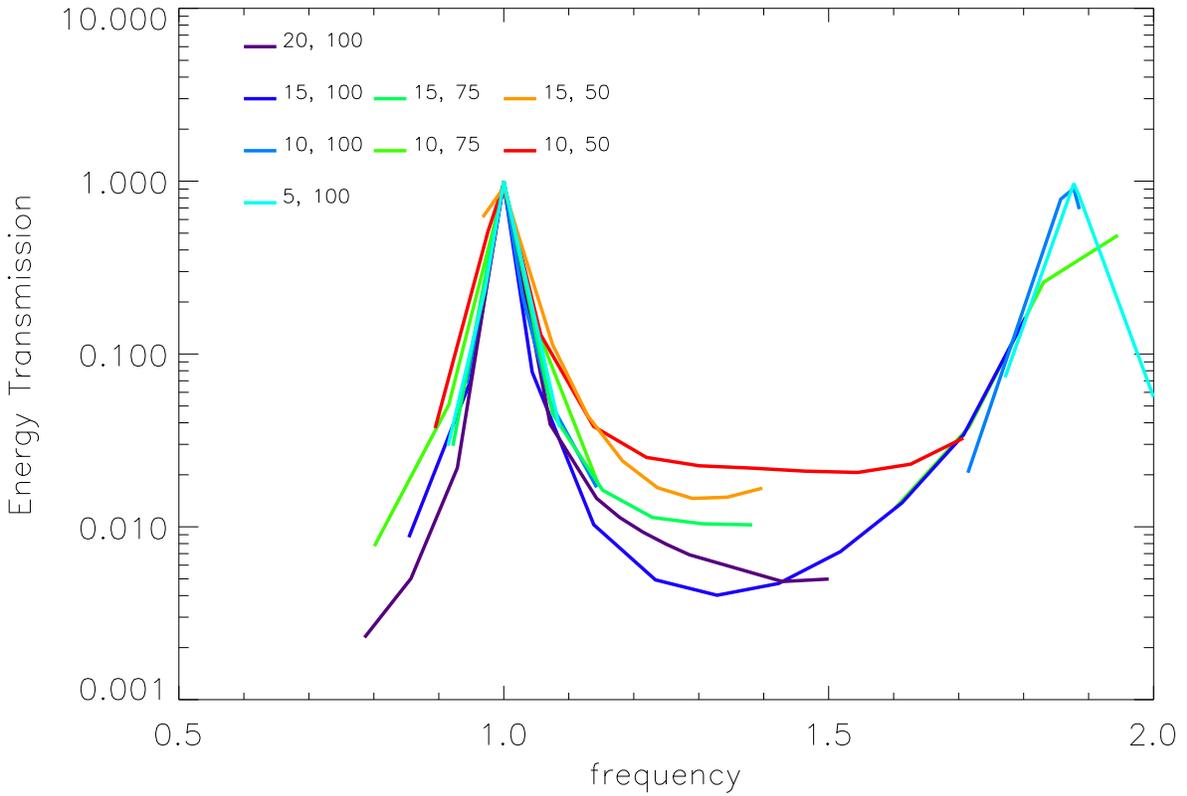}
\caption{The energy transmission coefficient as a function of wave frequency (expressed in terms
of the loop resonant frequency) for the loop lengths and magnetic fields considered in this paper.
A transmission coefficient of unity is achieved on resonance for all loops. Away from resonance, shorter loops (with higher frequency waves) transmit more energy between the chromosphere and the corona. The magnetic field variation is less significant.}
\end{figure}
\clearpage

\begin{figure}
\includegraphics[width=6.4in]{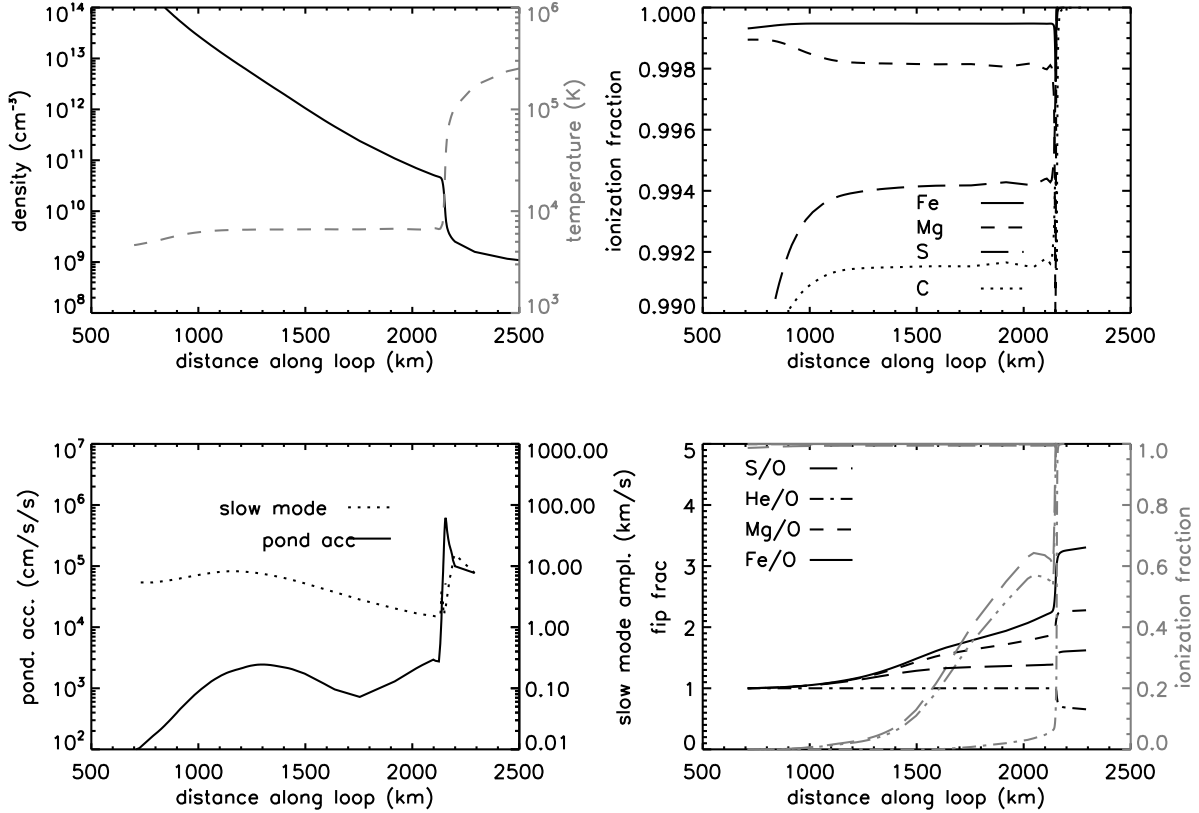}
\caption{Chromospheric section of the model loop, length 75,000 km, magnetic field
15 G, angular frequency 0.07 rad s$^{-1}$. The upper left panel gives the density (solid
line) and temperature (dashed line). Bottom left gives the  ponderomotive force
and slow mode wave (i.e. longitudinal wave excited by the Alfv\'en waves) amplitude.
Upper right; ionization fractions of Fe,
Mg, S and C. Bottom right; fractionations of Fe, Mg, S, and He, as well
as the ionization fractions in grey of H (dash), C (dash triple dot),
and He (dash dot). }
\end{figure}
\clearpage

\begin{figure}
\includegraphics[width=6.4in]{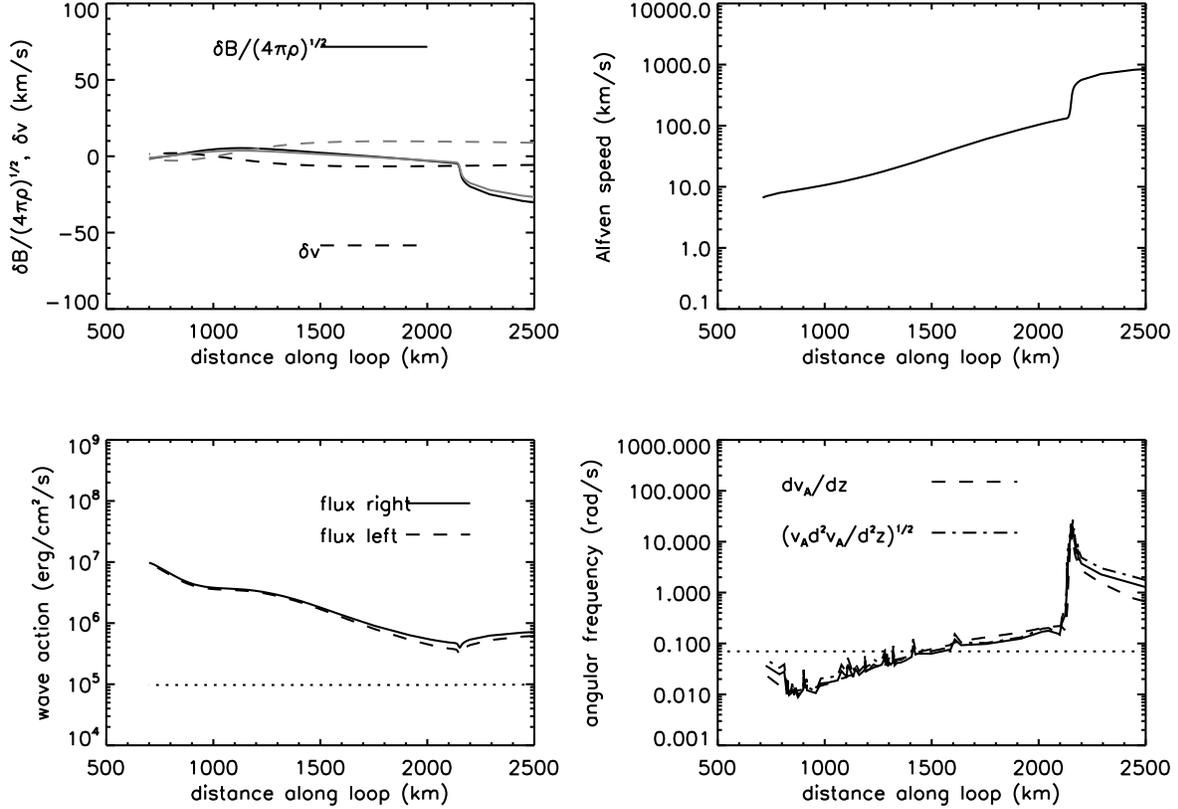}
\caption{Chromospheric section of the model loop, length 75,000 km, magnetic field
15 G, angular frequency 0.07 rad s$^{-1}$. The upper
left panel gives the amplitudes of Els\"asser variables in km s$^{-1}$ ($\delta B/\sqrt{4\pi\rho}$ solid lines, $\delta v$ dashed lines, real parts in black and imaginary parts in gray.)
At lower left are shown the wave energy fluxes in erg cm$^{-2}$s$^{-1}$ for left and right going waves. The thin dotted line shows the difference in energy fluxes divided by the magnetic
field strength and should be a horizontal line if energy is conserved. Upper right gives the
Alfv\'en speed $V_A$ profile through the chromosphere. The bottom right right gives the variation of
$\partial V_A/\partial z$ (dashed line) and $\sqrt{V_A\partial ^2V_A/\partial z^2}$ (dot-dashed line). The solid line gives $\sqrt{\left(\partial V_A/\partial z\right)^2/4+\left|V_A\partial ^2V_A/\partial z^2\right|/2}$, which is the Alfv\'en wave cutoff frequency as given by Moore et al. (1991) and
Musielak et al. (1992).}
\end{figure}
\clearpage

\begin{figure}
\includegraphics[width=6.4in]{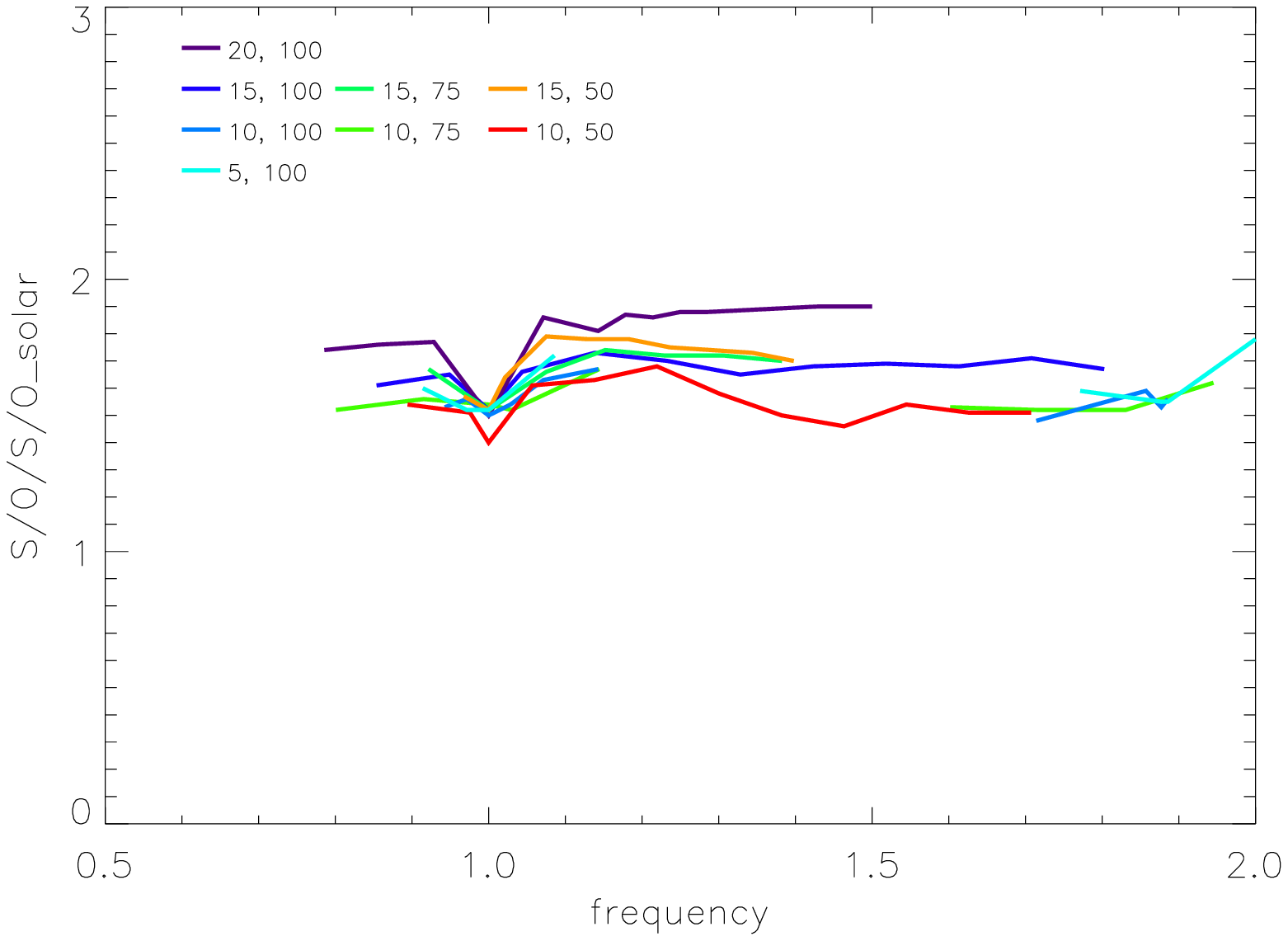}
\caption{S/O relative to their solar ratio as a function of the
frequency divided by the resonance frequency, similar to Figure 4.}
\end{figure}
\clearpage

\begin{figure}
\includegraphics[width=6.4in]{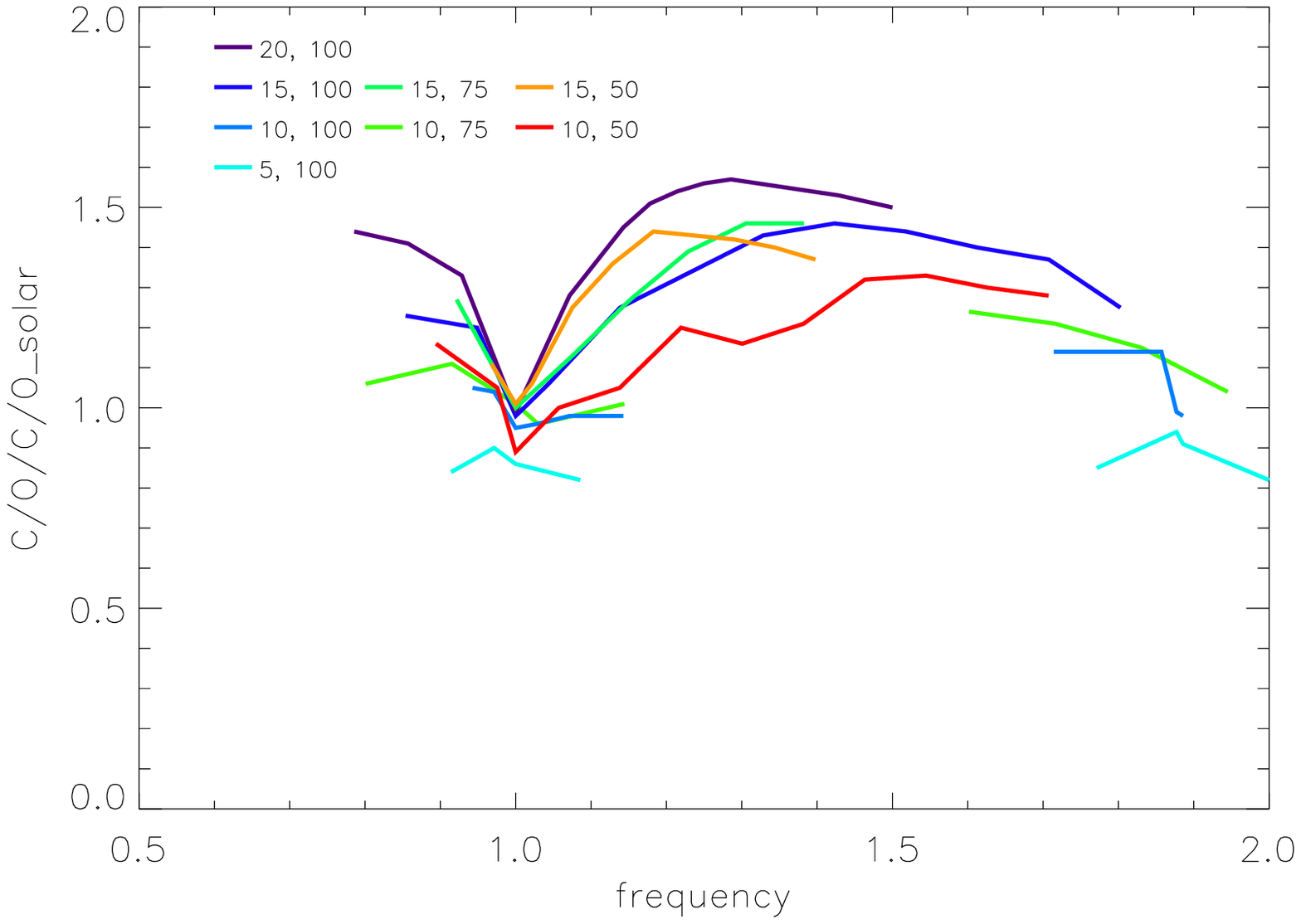}
\caption{C/O relative to their solar ratio as a function of the
frequency divided by the resonance frequency, similar to Figure 4.}
\end{figure}
\clearpage

\begin{figure}
\includegraphics[width=6.4in]{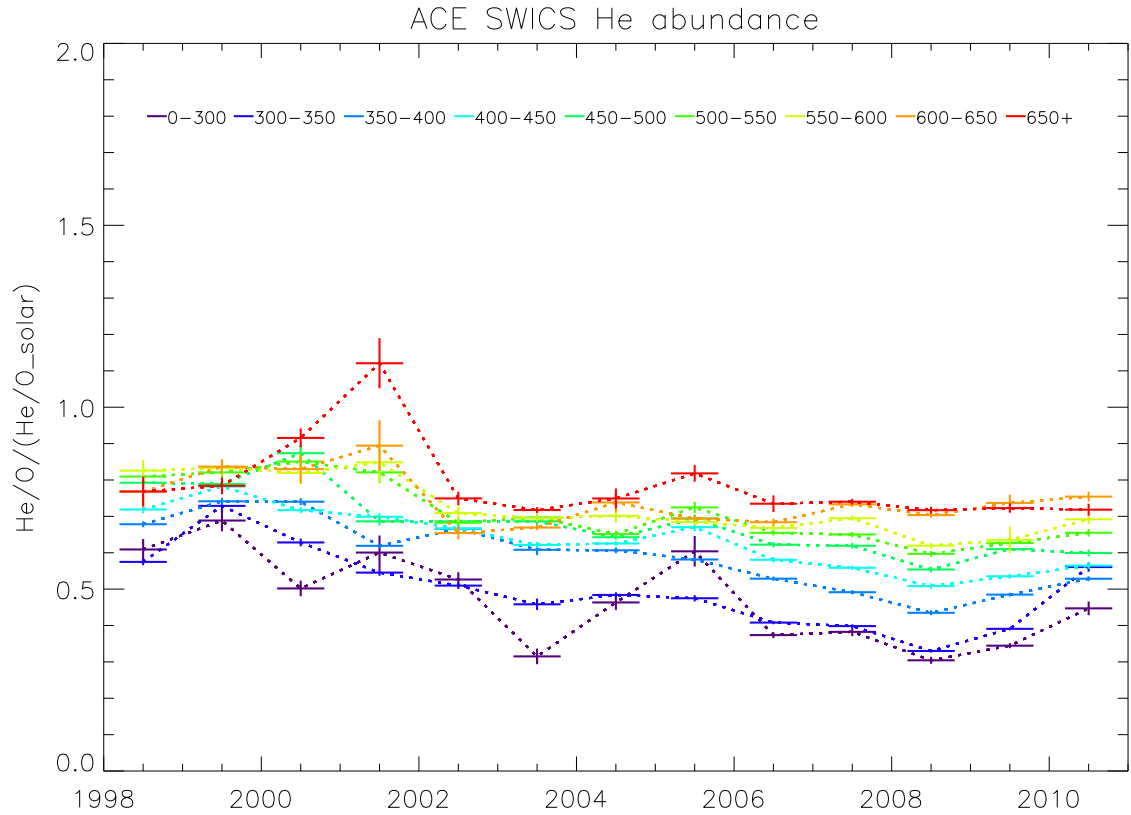}
\caption{ACE SWICS yearly average of the measured 2 hr averaged He/O ratio relative to
photospheric (126) from 1998 to 2010 as grouped by He velocity in 50
km~s$^{-1}$ bins. Errors are the standard error on the mean. A trend of higher He/O abundance
ratio with higher solar wind speed is clearly visible, though strong variation with phase of the
sunspot cycle is not seen.}
\end{figure}
\clearpage

\begin{figure}
\includegraphics[width=6.4in]{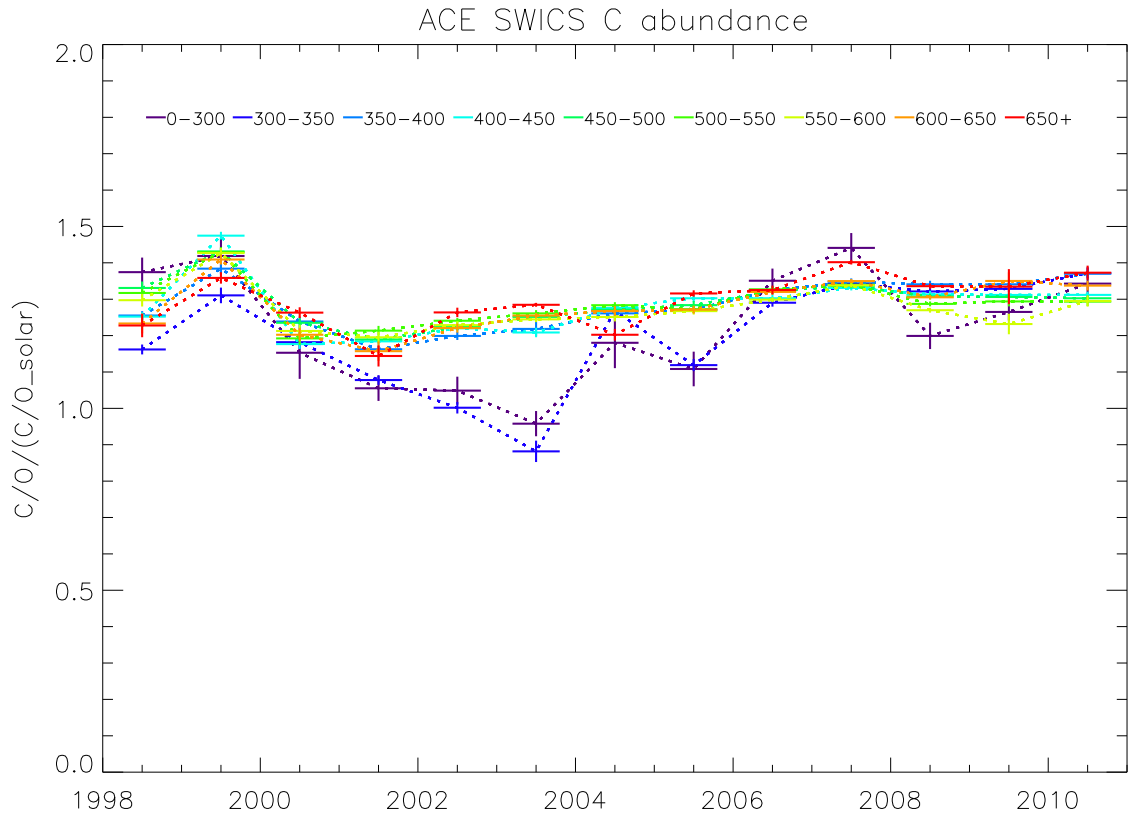}
\caption{Same as Figure 11, but for C/O relative to photospheric (0.550). A similar correlation of
C/O with wind speed is seen, especially around solar minimum years 2002 - 2004, and 2006.}
\end{figure}

\end{document}